\documentclass[aps, reprint, onecolumn, tightenlines, longbibliography, a4paper,
  showpacs, 11pt, eqsecnum]{revtex4-1}
\usepackage[english]{babel}

\usepackage[linktocpage=true]{hyperref} 

\hypersetup{
    colorlinks=true,
    linkcolor=blue,
    filecolor=blue,      
    urlcolor=blue,
}

\usepackage{color}

\newcounter{mnotecount}[section]

\renewcommand{\themnotecount}{\thesection.\arabic{mnotecount}}

\newcommand{\mnote}[1]{\protect{\stepcounter{mnotecount}}${\raisebox{0.5\baselineskip}[0pt]{\makebox[0pt][c]{\color{magenta}{\tiny\em$\bullet$\themnotecount}}}}$\marginpar{\raggedright\tiny\em$\!\bullet$\themnotecount:
#1}\ignorespaces}

\renewcommand{\mnote}[1]{}        %no Marginal Note

\usepackage{graphicx}
\usepackage{bm}% bold math

\usepackage{psfrag}

\usepackage{amssymb}

\usepackage{comment} 
\excludecomment{oldversion}

\newcommand{\bb}{{\ensuremath{\cal B}}}
\newcommand{\Gg}{{\ensuremath{\cal G}}}
\newcommand{\mm}{{\ensuremath{\cal M}}}

\newcommand{\cc}{{\ensuremath{\cal C}}}
\newcommand{\DD}{{\ensuremath{\cal D}}}
\newcommand{\AAA}{{\ensuremath{\cal A}}}
\newcommand{\dd}{\ensuremath{\textrm{d}}}
\newcommand{\ff}{{\ensuremath{\cal F}}}
\newcommand{\Lie}{{\ensuremath{\cal L}}}
\newcommand{\zz}{{\ensuremath{\cal Z}}}
\newcommand{\dyv}{\textrm{div }}
\newcommand{\const}{\textrm{const}}
\newcommand{\UD}[2]{\ensuremath{^{#1}_{\phantom{#1} #2}}}

\newcommand{\UDD}[3]{\ensuremath{^{#1}_{\phantom{#1} #2 #3}}}

\newcommand{\pard}[2]{\ensuremath{\frac{\partial #1}{\partial #2}}}
\newcommand{\RealNum}{\ensuremath{\textbf{R}}}

\newtheorem{theorem}{Theorem}[section]

\newtheorem{corollary}[theorem]{Corollary}

\newenvironment{definition}[1][Definition]{\begin{trivlist}
\item[\hskip \labelsep {\bfseries #1}]}{\end{trivlist}}

\newcommand{\qed}{\nobreak \ifvmode \relax \else
      \ifdim\lastskip<1.5em \hskip-\lastskip
      \hskip1.5em plus0em minus0.5em \fi \nobreak
      \vrule height0.75em width0.5em depth0.25em\fi}

\begin{document}

\preprint{}

\title{Variational principle for the Einstein--Vlasov equations}% Force line breaks with \\

\author{Lars Andersson}%
 \email{lars.andersson@aei.mpg.de}
\affiliation{Max--Planck--Institut f\"ur Gravitationsphysik (Albert--Einstein--Institut), Am M\"uhlenberg 1,
14471 Potsdam/Golm, Germany}

\author{Miko\l{}aj Korzy\'nski}
 \email{korzynski@cft.edu.pl}
% \email{mikolaj.korzynski@aei.mpg.de}%Lines break automatically or can be forced with \\
\affiliation{%
 Center for Theoretical Physics, Polish Academy of Sciences,
Al. Lotnik\'ow 32/46, 02-668 Warsaw, Poland
}%

%\date{\today \ {\tt File:\jobname{.tex}}}

\begin{abstract}
The Einstein--Vlasov equations govern Einstein spacetimes
filled with matter which interacts only via gravitation.
The matter, described by a distribution function on phase space, evolves under
the collisionless Boltzmann equation, corresponding to the free geodesic motion
of the particles, while the source of the gravitational field  is given by 
the stress--energy 
tensor defined in terms of momenta of the
distribution function. 
As no variational derivation of the
Einstein--Vlasov system appears to exist in the literature, 
we here set out to fill this gap. In our approach we treat the matter as a generalized type of fluid, flowing in the tangent bundle instead of the spacetime. We present the actions for the Einstein--Vlasov system in both the Lagrangian and Eulerian pictures.
\end{abstract}

\maketitle

%temporarily use toc
\tableofcontents

\section{Introduction} 
In this paper we consider general relativistic spacetimes filled with a
system of particles 
%in a disordered state of motion, 
interacting only via gravitation. 
The system of particles may be represented by a
distribution function on phase space. 
%
%A system of particles in a disordered state of motion, interacting only via
%their gravitational field is described in general relativity by the
%Einstein-Vlasov equation. 
%
%The state of motion of the particles can be
%represented in terms of a distribution function on phase space supported on
%the bundle of future causal vectors. \mnote{timelike? co-vectors?}  
%
%In general relativity, the geometry of a spacetime 
%%$(\mm, g_{ab})$ 
%filled with particles interacting only via 
%%collisions which conserve energy-momentum
%their 
%gravitational field is governed by the Einstein--Vlasov equation. 
%%is governed by an Einstein-Boltzmann equation. 
By the postulates of general relativity, 
free particles move along future directed 
geodesics and hence the distribution function
evolves by flowing along the geodesic spray, i.e. the vector field on phase
space which generates the geodesic flow. 
%The world line of a freely moving particle is a causal geodesic and hence the
%system of particles can be described by a
%distribution function 
%$f$ 
%on phase
%space, supported in the bundle of future solid light cones and 
%evolving under flow of the geodesic spray. 
This flow equation is known as the
the Vlasov equation,
also known as the collisionless Boltzmann or Liouville equation.  
\mnote{or co-vectors?}
The Einstein equation couples 
matter to the gravitational field via the stress energy tensor, the
components of which are 
defined in terms of phase space momenta of the distribution
function. The resulting system of field equations is known as the
Einstein-Vlasov system. If we consider systems of particles which
self-interact via collisions, the Vlasov equation must be replaced by a
Boltzmann equation with collision term and we are led along similar lines to
considering an Einstein-Boltzmann system.

%
%The macroscopic properties of the matter system expressed in terms
%of the components of the stress energy tensor thus represent
%macroscopic properties of the matter system defined from the microscopic
%state of the matter system. 

Matter models described by a distribution function on phase space, 
evolving under some version of the Boltzmann equation are known as kinetic
matter models. Since each spacetime volume element may contain particles
moving in all directions, the matter is in a disordered state of motion. On
the other hand, in a continuum matter model such as a perfect
fluid, which may be viewed as a degenerate limit of kinetic models, 
the system of particles is in an ordered state of motion,  
where all particles at a given spacetime point 
move along the fluid velocity field. 
%in the same direction. 
The state of a fluid, and its equations of motion are given directly in terms
of its macroscopic properties, eg. the components of the stress energy
tensor. On the other hand, the distribution function of a kinetic matter
model provides a microscopic representation of the matter from which its
macroscopic properties may be calculated. 
\mnote{also mention connection to statistical mechanics, see Kandrup-Israel
  \cite{israel:kandrup:1984AnPhy.152...30I,kandrup:1984PhyA..126..461K} see
  also Tian \cite{tian:2009AnP....18..783T} ?? -- also cf BBGKY hierarchy
  etc.}
\mnote{also mention Bernstein \cite{bernstein:1988kteu.book.....B} --
  connection to structure formation etc. cf. 
\cite{tsagas:challinor:maartens:2008PhR...465...61T}} 
Kinetic matter models including those where the matter particles couple 
to fundamental fields, such as the
Vlasov-Maxwell, Vlasov-Poisson and the Einstein-Vlasov systems and their
collisional counterparts play an
important role over a vast range of scales and the literature is huge. We
refer to the books
\cite{degroot:vanleeuwen:vanweert:MR635279,stewart:1971LNP....10.....S} and
the survey papers \cite{andreasson:2011LRR....14....4A,rein:2007:MR2549372}
for references. 

General relativistic kinetic theory was first considered in the 1930's by 
Synge \cite{synge:1934} and 
Walker \cite{walker:1936}, see however the brief discussion by Eddington
\cite[p. 116]{eddington:1924:MR0115743}.  
In his 1957 book \cite{synge:1957}, Synge
presented the theory of gases in special relativity in a nice geometric
form. This formulation influenced several discussion of the foundations of
kinetic theory in general relativity, both for collisionless (Vlasov) matter
as well as matter governed by the Boltzmann equation, including the papers of 
Tauber and Weinberger \cite{tauber:weinberger:1961}, Chernikov 
\cite{chernikov:1962SPhD....7..414C, chernikov:1962SPhD....7..397C,
  chernikov:1962SPhD....7..428C}, Lindquist
\cite{lindquist:1966}. Ehlers 
wrote several survey articles on the subject\footnote{Presumably based on his lectures on kinetic theory and general
relativity at U. Texas 1964-65 (unpublished, cf. \cite{lindquist:1966}).} 
including 
\cite{ehlers:kinetic:1971}.  
For more recent surveys, see the articles by  Rendall \cite{rendall:1997} and
Andr\'easson \cite{andreasson:2011LRR....14....4A}.
%
%In order to study the dynamics of the Einstein-Vlasov system it is necessary
%to have a well-posed initial value problem. 
Local existence of solutions to
the Cauchy problem for the 
Einstein-Vlasov system has been proved by 
\mnote{what about uniqueness, global uniqueness?} 
Choquet-Bruhat \cite{choquet-bruhat:1971:MR0337248}, see also
\cite{bancel:choquet-bruhat:1973:MR0356790} for results on the
Einstein-Boltzmann equations. For the Einstein-Boltzmann case, the results
appear rather limited. A recent treatment of the Cauchy problem for the
Einstein-Vlasov equations has been given by Ringstrom
\cite{ringstrom:EV:book}. The non-linear stability of Minkowski space in Einstein-Vlasov  theory is known for both the massless and the massive case, cf. \cite{MR3629140, 2019arXiv190312251J,2017arXiv170706141F,2019ArRMA.tmp..102L}

The Einstein-Vlasov system can be expected to be relevant for understanding
the evolution of cosmological models, and large scale structure formation in
the universe. However, due to the difficulties posed, most 
work on global
properties of the Einstein-Vlasov system has been done under strong symmetry
assumptions. For numerical studies see eg. 
\cite{shapiro:teukolsky:1986ApJ...307..575S,olabarrieta:choptuik:2002PhRvD..65b4007O,andreasson:rein:2006CQGra..23.3659A}. Using
analytical methods, 
issues of dynamical stability and singularity formation for the
Einstein-Vlasov system have been studied in highly symmetric situations, see 
\cite{andreasson:2011LRR....14....4A} for further references. Fajman
\cite{fajman:2011} has proved dynamical stability of 
certain  Einstein-Vlasov 
cosmological models with $\text{\rm U}(1)$ symmetry. Future stability for the Milne model in Einstein-Vlasov theory has been proved \cite{2017arXiv170900267A}. 
However, 
little is known in general. An exception is the work of Ringstr\"om on the
stability of cosmological models with positive cosmological constant
containing collisionless matter \cite{ringstrom:EV:book}, which is not
dependent on any symmetry assumption. 
In contrast to the
Einstein-Vlasov case, large scale N-body simulations in Newtonian gravity 
have been performed without symmetry assumptions as part of studies of large
scale structure formation. The N-body dynamics is related to the
Vlasov-Poisson system, 
see eg. \cite{joyce:2008CNSNS..13..100J} and
references therein for a discussion of the relation of these studies to the
Vlasov-Poisson system, see also \cite{rein:2007:MR2549372}. 
\mnote{mention Coles et al on wave approx??} 
The question of stability of steady states for 
kinetic matter models has been
widely studied in astrophysics and cosmology, mainly in the case of Newtonian
gravity see
eg. \cite{rein:2007:MR2549372, binney:tremaine:2008gady.book.....B}.
A few references on the Einstein-Vlasov system which are relevant here are 
\cite{ipser:thorne:1968, 
  fackerell:1968ApJ...153..643F, ipser:1980ApJ...238.1101I, kandrup:oneill:1994}. 
%These papers deal
%mainly with issues relating to the stability of steady states. 
Whereas a number of theorems on providing conditions under which stability of
steady states holds for the Vlasov-Poisson system, 
the picture is not as complete in the general relativistic case, see however 
\cite{2018arXiv181000809H}.
%Wolansky \cite{wolansky:2001} and references therein. 
In studies of
stability, the energy methods, including the energy-Casimir method,  
play an important role. Hamiltonian formulations
for the Einstein-Vlasov system
has been discussed by Kandrup and collaborators, see 
\cite{kandrup:1991NYASA.631...88K,kandrup:oneill:1994,kandrup:morrison:1993AnPhy.225..114K}. 

In view of the above, it appears natural to make a systematic study of
variational formulations of the Einstein-Vlasov and related systems. However, 
although there is a considerable literature on variational formulations of
eg. Vlasov-Maxwell and related theories, see
\cite{ye:morrison:1992PhFlB...4..771Y,morrison:2005PhPl...12e8102M},  apart
from the above mentioned work by Kandrup et al. on the Hamiltonian for
Einstein-Vlasov, there appears to be no such treatment in the literature, see
however \cite{okabe:morrison:etal:2011PhRvD..84b4011O} for recent work on a 
variational formulation for Einstein-Vlasov in the spatially homogeneous
case.

%\section{OLD Introduction}

\begin{oldversion}
The Einstein--Vlasov equations describe spacetimes with a source for the 
gravitational field of the form of a collection of particles. In the standard formulation of the
Vlasov matter the individual particles constituting the matter are described from a \emph{microscopic} point of view,
in terms of a distribution function in the single--particle phase space. Thus at a given point
of spacetime we consider separately the motion all particles that pass through it. 
This is in contrast to more usual \emph{macroscopic} matter models, like dust, fluids or elastic matter, where the state of 
matter at a given point of the spacetime is described 
from the point of view  of statistical mechanics, via the local density, local velocity and an equation of state.
\end{oldversion}

Let $\mm$ be the spacetime equipped with a metric tensor $g$ and a coordinate system 
$x^\mu$ with $x^0$ being the time variable.
Recall that the distribution function is a positive function on the cotangent bundle, whose support is contained within the 
future pointing null cones. With the help of the coordinate system it can be written as 
$f(x^\mu, p_\nu)$, depending on particle position and four--momentum. Up to a coefficient, $f$ gives
the number of particles contained within a volume element in the spacetime and whose four--momentum is close
to $p_\mu$, as measured by an observer at that point. The precise definition of $f$ will be given later in the paper,
see also \cite{andreasson:2011LRR....14....4A}.

Function $f$ evolves under the relativistic, collisionless Boltzmann equation, 
which corresponds to the assumption
that the individual matter particles move along geodesics.
The equation is known in this context as the the Liouville or Vlasov equation
\begin{eqnarray}
p^\mu\,\pard{f}{x^\mu} + p^\nu\,p_\mu\,\Gamma\UDD{\mu}{\nu}{\sigma}\,\pard{f}{p_\sigma} = 0 \label{eqVlasov1}.
\end{eqnarray}
Note that in 
contrast to dust, fluids or
elastic sources, we assume here that infinitely many particles may pass through each point of spacetime, distinguished only by the values
of their four--momenta.

The gravitational field  is coupled to the matter content via Einstein equations of the form
\begin{eqnarray}
 &&R^{\mu\nu} - \frac{1}{2}\,R\,g^{\mu\nu} = 8\pi G\,T^{\mu\nu} \label{eqQQQ2} \\ 
 &&T^{\mu\nu} = (-g)^{-1/2}\, \int \dd^4 p_\mu \,f(x^\mu,p_\nu)\,\frac{p^\mu\,p^\nu}{m} \label{eqEinsteinVlasov1},
\end{eqnarray}
where $m = \sqrt{-p_\alpha\,p^\alpha}$. 
The stress--energy tensor at a given point is simply the integral over all possible four--momenta with each particle
contributing its four--momentum to the total energy and momentum densities and stress. 

For early discussions of collisionless matter in general relativity, see
 \cite{tauber:weinberger:1961,lindquist:1966,ehlers:1973,ehlers:kinetic:1971,ipser:thorne:1968} and references
therein. In these works, the collisionless matter is viewed as arising from a statistical treatment of e.g. a gravitationally interacting gas
\cite{tauber:weinberger:1961,lindquist:1966}, stars or
galaxies \cite{ipser:thorne:1968}.  
From this point of view, the stability of steady states of
self-gravitating collisionless matter is of importance, and has been
considered in numerous papers, see eg. 
\cite{ipser:thorne:1968,kandrup:oneill:1994}.

So far equations (\ref{eqVlasov1})--(\ref{eqEinsteinVlasov1}) have been derived via phenomenological considerations. 
In this paper we will show how to arrive to them using
a variational principle. We will show that they can be thought of as a consequence of vanishing of variation of action $S$ of the form
\begin{eqnarray}
 S = \frac{1}{16\pi G} \int_{\AAA} R\, \sqrt{-g}\, \dd^4 x + S_{m}[g_{\mu\nu}, \Phi^a] 
\end{eqnarray}
where ${\AAA}$ is a compact domain in $\mm$.
 The first term is the standard functional depending on the metric tensor (and connection field if we consider a
 Palatini--type variational principle), while the second, matter term, is a functional depending on
both the metric and the matter degrees of freedom, represented here by $\Phi^a$. 
It is a standard result that vanishing of variation with respect to the metric yields Einstein equations of the form
(\ref{eqQQQ2})
with
\begin{eqnarray}
 T^{\mu\nu} = \frac{2}{\sqrt{-g}}\, \frac{\delta S_m}{\delta g_{\mu\nu}}. \label{eqQQQ1}
\end{eqnarray}
Variations with respect to $\Phi^a$ on the other hand should give the Vlasov equation (\ref{eqVlasov1}). 
Let us emphasize here that it is crucial that the variational principle gives not only (\ref{eqEinsteinVlasov1}), but also 
(\ref{eqVlasov1}). This in contrast to macroscopic matter models like dust or fluids, where
the matter equations follow from the conservation of the stress--energy tensor (\ref{eqQQQ1}) 
appearing in (\ref{eqEinsteinVlasov1}). We can therefore forget about variations 
with respect to matter degrees of freedom 
altogether, since 
the matter equations of motion can be derived indirectly from variations with respect to the metric, giving
the expression for $T^{\mu\nu}$, and the Bianchi 
identities, which imply that $T^{\mu\nu}$ is conserved \cite{kijowski-1998}. 
This is not the case in Einstein--Vlasov equations, 
where (\ref{eqEinsteinVlasov1}) 
does not imply (\ref{eqVlasov1}).

The variational formulation of the 
Einstein equations with macroscopic
matter has been studied by  many authors. In the literature one can
clearly distinguish two main approaches to the problem, differing with
respect to how the matter degrees of freedom  are treated.   

Let $\bb$ denote the material space, i.e. collection of particles or fluid elements, endowed with the structure
of a three--dimensional differentiable manifold and other geometric objects describing the internal state of matter
(see \cite{KIJOWSKI1992207, PhysRevD.41.1875, kijowski-1998, kijowski-book-2005}). 
One can now describe the  matter degrees of freedom in terms of \emph{configurations}, i.e. mappings
$C: \mm \to \bb$ which assign to every point of the spacetime the particle which passes through it.
Mappings $C$ can be used to define the local particle current $n^\mu$ on $\mm$, a vector density describing the motion of the 
particles through the spacetime and at the same time measuring their number in a small volume of $\mm$.
$n^\mu$ can in turn be broken into the relativistic particle density and the four--velocity 
\begin{eqnarray}
 n^\mu &=& \rho\,u^\mu.
\end{eqnarray}
 The current satisfies a
conservation equation
\begin{eqnarray}
 \frac{\partial n^\mu}{\partial x^\mu} = 0,
\end{eqnarray}
 which corresponds
to the fact that the particles themselves are neither destroyed nor created during the motion. It is a kinematical
conservation law, satisfied by the virtue of the definition of $n^\mu$, irrespective of the equations of motion.
It is also possible to consider $n^\mu$ rather than $C$ as the field which encodes the state of matter.
This way of describing the macroscopic matter is called the Eulerian picture.

Altenatively one might use the Lagrangian picture, where the direction of mappings is reversed. 
Consider a foliation of $\mm$ by three--dimensional, spacelike constant time slices $\mm_t$. One can invert at least localy the mapping
restriction of $C$  to $\mm_t$.  We can thus describe the state of matter by  mappings $D_t: \bb \to \mm_t$, called 
\emph{deformations}.
For a fixed $p \in \bb$ and variable $t$ the deformations $D_t$ give exactly the worldline of the particle $p$.  
With the help of a coordinate system on $\mm$ the
mapping $D_t$ can be expressed via a triple of scalar functions (fields) on
$\bb$.

Both field theoretic descriptions can be used to define variational principles for the Einstein equations
with matter. In the Eulerian picture one encounters a small complication
due to the fact that we should not consider all variations of the particle current
field,  but only those which arise from displacing individual world--lines of
particles, without altering their number. This kind of constrained variational
principle was pioneered by Taub \cite{taub-1954} (see  also
\cite{carter-fluid}). From the variational principle we obtain automatically the conserved stress--energy tensor of the
matter, which serves as the gravitational field source in the Einstein equations.

In this paper we will demonstrate that it is possible to describe the Vlasov
matter using both Lagrangian and Eulerian variables and derive the
Einstein--Vlasov  equations using variational principles similar to
the ones used in macroscopic matter. 
The main difference lies in the domain and target spaces
of mappings describing the state of matter: the matter space $\bb$ must be assumed to be 7--dimensional,
 since particles
in the microscopic matter models are labeled by their position in the spacetime as well as their four--momentum, while the 
target space must be the phase space of massive particles, which is isomorphic to the cotangent bundle
of the spacetime. 

Despite technical differences, both Eulerian and Lagrangian pictures of the Vlasov matter are
quite similar to their macroscopic counterparts. Namely, the configurations are the mappings from the
union of future null cones in the cotangent bundle $P^*_+$ to $\bb$
\begin{eqnarray}
 \cc : P^*_+ \to \bb.
\end{eqnarray}
From the kinematical point of view $P^*_+$ is merely the phase space of a massive particle in $M$.
Just like in the macroscopic matter models, $\cc$ gives rise to a conserved matter current vector density $J$ on $P^*_+$. It is a geometric object
defined on $P^*_+$ and cannot be projected down to 
the spacetime $\mm$, but just like in the macroscopic case it can be broken into
the invariant particle density and the normalized velocity vector $U$, both defined on $P^*_+$. The invariant density is 
directly related to the distribution function $f$. 

The conserved matter current on $P^*_+$ can be integrated partially over all momenta in each point of the spacetime. This 
gives rise to a conserved particle current on the spacetime, in the form of a densitized four--vector field. 
In the same way $J$ combined with mass can give rise to the mass current on the spacetime. 

The matter part of the action is given in the form of a functional $S_m$ of $J$. The vanishing of its 
constrained variations with respect to $J$ implies
that $U$ is the geodesic spray vector field $G$, generating lifts of the geodesics to $T^*\mm$.
This implies that the individual particles of the Vlasov matter follow geodesics in their motion.
The Vlasov equation (\ref{eqVlasov1}) for $f$ is a simple consequence of that fact and of the conservation law for $J$.
The Einstein equation with the appropriate stress--energy tensor follows from varying of $S_m$ with respect to
the metric.

The Lagrangian picture is even simpler. It is possible to consider deformations,
defined as mappings from $\bb$ to the constant time slice of $P^*_+$
\begin{eqnarray}
 \DD_t : \bb \to P^*_{+,t}
\end{eqnarray}
as the set of variables describing the state of the matter degrees of freedom. As before, $\DD_t$ is the 
inverse of $\cc$ at a given instant of coordinate time. It turns out however that the picture can be simplified if
instead of $\DD_t$ we use its projection $\widetilde\DD_t$ down to $\mm$
\begin{eqnarray}
 \widetilde\DD_t &:& \bb \to \mm_t \\
 \widetilde\DD_t &=& \Pi\circ\DD_t,
\end{eqnarray}
where $\Pi$ is the canonical projection of the bundle $P^*_+$
 The variational principle for the matter can be expressed
in terms of $\widetilde\DD_t$ and its time derivative. 
With the help of a coordinate system on $\mm$ we may encode $\widetilde\DD_t$ in three fields on $\bb$ corresponding
to positions of particles at a given moment. This way we reduce down to 3 the number of fields required to describe 
the state of the system. The Lagrangian density of the action turns out to be 
essentially the integral of homogeneous single--particle
Lagrangians over the material space.

\subsection{Overview of this paper}
%The paper is organized as follows. 
In section \ref{sec:prel} 
we discuss the most
important geometric notions and the notation we use throughout the paper. In
section \ref{sec:standard-approach} we recall 
the Einstein--Vlasov equations in the standard
formalism and then present a new one, inspired by Kijowski and Magli articles
on relativistic elastomechanics. Finally in the last two sections we present
two formulations of variational principles leading to Einstein--Vlasov
equations: one emplying the Lagrangian approach to the matter and the other
using the Eulerian one. We also derive the Hamiltionian of the
Einstein--Vlasov system system. In the appendix we review the most important
properties of vector densities and prove a technical result needed in section
\ref{sec:eul}.
\mnote{LA:put in ref for section VII -- check}

\section{Preliminaries and notation}\label{sec:prel} 
In this section we will present briefly the most important geometric concepts and clarify the notation we will use throughout the work.
%For a more detailed introduction see for example \cite{ehlers:1973, andreasson:2011LRR....14....4A}.

Let $(\mm,g)$ be  a time--oriented, compact pseudo--Riemannian manifold of dimension 4 equipped with a Lorentzian metric
$g$ of signature $(-,+,+,+)$, topologically a Cartesian product
of an interval $I = \left[t_0,t_1\right]$ and a three--dimensional manifold $S$.
We introduce a coordinate system $(x^\mu)$ on $\mm$, consistent with the orientation, whose first coordinate $x^0$ is
timelike and future oriented. By $S_{t}$ we shall denote the (spacelike) constant coordinate time surface in $\mm$, homeomorphic to $S$.

The tangent and cotangent spaces over a point $p\in \mm$ are collections of all vectors or covectors at that point. Given the coordinate system on $\mm$
they can be parametrized by the four components in the decomposition in the natural bases $\pard{}{x^\mu}$ or $\dd x^\mu$ respectively. Namely, in the tangent bundle we have coordinates $\zeta^\mu$ given by equation 
$X_p = \zeta^\mu\,\frac{\partial}{\partial x^\mu}$ where $X_p$ is a vector tangent at $p$, and by analogy in the cotangent we have $v_\mu$ defined by
$\omega_p = v_\mu\,\dd x^\mu$, where $\omega_p$ is a covector at $p$. The unions of all tangent or cotangent spaces are manifolds themselves, the  tangent $T\mm$ and cotangent $T^*\mm$ bundles respectively. The construction given above, together with $(x^\mu)$, yields coordinate systems $(x^\mu, \xi^\nu)$ and $(x^\mu, v_\nu)$ in the two aforementioned bundles. 

The metric tensor gives a natural isomorphism between the tangent and cotangent spaces by lowering indices of vector fields
\begin{eqnarray}
   \alpha_p : T_p\mm \ni X_p \mapsto g_p(X_p,\cdot) \in T^*_p\mm.
\end{eqnarray}
This pointwise isomorphism of spaces extends to the isomorphism of the whole bundles
$\alpha : T\mm \mapsto T^*\mm$.
Note that $\alpha$ allows us to map or push forward any construction made in $T\mm$ to $T^*\mm$
and vice versa. In this paper we will be more concerned with the latter, we will therefore recall some facts from its geometry.

In a manifold with a non--degenerate metric the cotangent spaces at each point are endowed with natural volume forms given by $\eta = |-g|^{-1/2}\,\dd v_0\wedge\dd v_1 \wedge\dd v_2 \wedge \dd v_3$. 
The whole cotangent bundle has a preferred volume form given by the product of the natural volume $\rho$ on $\mm$ ($\rho =|-g|^{1/2}\,\dd x^0\wedge
\dd x^1\wedge\dd x^2\wedge\dd x^3$), and $\eta$
\begin{eqnarray}
 \kappa = \rho\wedge\eta = \dd x^0\wedge\dots\wedge\dd x^3\wedge \dd v_0\wedge\dots\wedge\dd v_3.
\end{eqnarray}
Moreover $T^*\mm$ has the structure of a symplectic manifold with the 2--form $\omega = \dd v_\mu\wedge \dd x^\mu$ (it can be checked that this definition does not depend on the coordinate system $x^\mu$).

The four--momenta of massive particles crossing a  point $x\in \mm$ are vectors and can thus be 
considered elements of $T_x\mm$. The four--momenta of massive particles are always spacelike and 
future pointing, so they always lie inside the future light cone. We will denote the bundle of future light cones
in $T\mm$ and $T^*\mm$ by $P_+$ and $P^*_+$ respectively.
$\alpha$ maps $P_+$ onto $P^*_+$ bijectively and $P^*_+$ inherits the volume form $\kappa$ from the cotangent bundle. 
The restriction of $P^*_+$  to constant coordinate time will be denoted by $\Sigma_t$, i.e.
\begin{eqnarray}
 \Sigma_t = \{(x,v)\in P^*_+ | x^0 = t \}.
\end{eqnarray}

The world--lines of massive particles constitute a special subclass of curves in the spacetime:
\begin{definition}
 The \emph{particle world--line} is a smooth, timelike curve parametrized by any function which monotonically increases with time. 
\end{definition}
In particular, it is possible to parametrize a world--line using the coordinate time $x^0$. 

Curves $x^\mu(s)$ with a given parametrization have unique lifts to curves on the cotangent tangent bundle, in which the vertical coordinates are given by the curve's tangent vector
\begin{eqnarray}
 \zeta^\mu = \dot x^\mu.
\end{eqnarray}
The curve may now be mapped by $\alpha$ to $T^*\mm$, which gives the lift of $x^\mu(s)$ to the cotangent bundle.
We will refer
to lifts of this kind as \emph{ordinary lifts}, opposed to lifts based on four--momenta of particle world--lines, which we will define below.
\begin{definition}
 The \emph{four--momentum lift} of the world--line $x^\mu(s)$ of a particle of mass $m > 0$ to $P_+^*$ is the curve in $P^*_+$ given by equation
 \begin{eqnarray}
  v_\mu(s) = \frac{m\,\dot x^\mu\,g_{\mu\nu}}{\sqrt{-\dot x^\alpha\,\dot x^\beta\,g_{\alpha\beta}}} \label{eq4mlift}
 \end{eqnarray}
(the $x^\mu(s)$ remains the same).
\end{definition}
This lift differs from the ordinary one in that the length of $v_\mu(s)$ is adjusted to the particle mass $m$ rather than the parametrization of the curve. 
Consequently $v_\mu$ is proportional to the tangent vector, but with a factor which ensures that $v_\mu$ is indeed equal to the four--momentum of the particle. 

From that definition we immediately infer the following
\begin{corollary}
A curve $(x^\mu(s), v_\nu(s))$ in $P^*_+$ is a four--momentum lift of a massive particle world--line if and only if its tangent vector $X = (\dot x^\mu(s), \dot v_\nu(s))$ satisfies
at each point the following equations:
\begin{enumerate}
 \item $\frac{\dd }{\dd s} (v_\mu(s)\,v_\nu(s)\,g^{\mu\nu}) = 0$ (the curve is confined to a single mass shell $v_\mu\,v_\nu\,g^{\mu\nu} = \const$) 
 \item $v_\mu(s) = C\,\dot x^\nu(s)\,g_{\nu\rho}$ for a $C \neq 0$ (vertical coordinates are proportional to the derivative of the horizontal ones). 
\end{enumerate}
\end{corollary}
Let $X$  be a nowhere vanishing vector field on $P^*_+$
\begin{eqnarray}
 X = X_x^\mu\,\pard{}{x^\mu} + X_{v\,\nu}\,\pard{}{v_\nu}
\end{eqnarray}
(lower case subscripts $x$ and $v$ will be our standard notation for the horizontal and vertical components of geometric objects on the bundle $P_+^*$).
$X$ generates a congruence of curves in $P_+^*$. The corollary above gives us the conditions under which all curves in congruence are particle world--line lifts, namely 
the vector field must be everywhere tangent to the mass shells and its horizontal component must be proportional to $v_\mu$
\begin{eqnarray}
  X\,(v_\mu\,v_\nu\,g^{\mu\nu}) &=& 0 \label{eqXvveq0} \\
  X_x^\mu &=& C\, v_\nu\,g^{\mu\nu} \qquad C > 0 \label{eqXxeqCv}.
\end{eqnarray}

The tangent and cotangent bundles of a (pseudo-)Riemannian manifold are automatically endowed with a special vector field called the \emph{geodesic spray}
\begin{eqnarray}
 G = g^{\mu\nu}\,v_\mu\,\pard{}{x^\nu} +
 \Gamma\UD{\mu}{\rho\sigma}\,g^{\rho\nu}\,v_\mu\,v_\nu\,\pard{}{v_\sigma} \label{eqGfield},
\end{eqnarray}
which generates the congruence of ordinary lifts of affinely parametrized geodesics. On $T^*\mm$ and $P_+^*$ it has the form given in (\ref{eqGfield}).
It can be checked that the integration curves of $G$ are at the same time the four--momentum lifts of geodesics, so $G$ satisfies conditions
(\ref{eqXvveq0}) and (\ref{eqXxeqCv}), and that it preserves the volume form $\kappa$
\begin{eqnarray}
 \Lie_G \kappa = 0
\end{eqnarray}
as well as the length of $v^\mu$
\begin{eqnarray}
 \Lie_G \sqrt{-v^\mu\,v_\mu} = 0.
\end{eqnarray}
With the help of $G$ equation (\ref{eqVlasov1}) takes a simpler form:
\begin{eqnarray}
 G(f) = 0 \label{eqGf}.
\end{eqnarray}

%\subsection{Distribution function and the Vlasov equation}

%The evolution of the system can be described by giving the metric on the manifold and the motion of all particles involved. Since in the Vlasov system we assume %that the particles of different four--momenta may occupy the same postion in the spacetime, and thus  it is sensible to consider the lifts of all particle's %world--lines to
%$P_+^*$. In $P_+^*$ each of the particles can be distinguished and no two trajectories intersect. 

\section{Standard approach to Einstein--Vlasov
  matter}\label{sec:standard-approach}

The physical interpretation of the distribution function $f$ at a given point can be spelled out in the following way:
let $u^\alpha$ be the four--velocity of an observer at $x^\mu$. The contraction $u\lrcorner \rho = V_u$, where $\rho$ is
the invariant 4--volume element on $\mm$, is an invariant 3--volume element on the subspace 
of $T_p\mm$ orthogonal to $u$ (and thus on any spacelike surface intersecting $p$ and orthogonal to $u$ at that point). 
The four--form  $\eta = (-g)^{-1/2} \dd v_0\wedge\cdots\wedge\dd v_3$ is the geometrically invariant volume form on
 the cotangent bundle and on $P_+^*$. The number of particles contained within $d_u V$ and
whose four momentum $v^\mu$ lies within $\eta$ is given by
\begin{eqnarray}
  dN = -f(x^\mu, v_\nu) \,\frac{v^\mu\,u_\mu}{m}\,d_u V\,\eta \label{eqfdef}
\end{eqnarray}
where $m = \sqrt{-v^\mu\,v_\mu}$. The additional factor $-\frac{v^\mu\,u_\mu}{m}$ ensures the geometric invariance of the definition
 (see \cite{andreasson:2011LRR....14....4A}).

Consider a spacelike hypersurface $W \in S_t$  and a domain $V$ in $\RealNum^4$. Let $N_{W\times V}$ denote the
number of particle world--lines intersecting $W$, for which the four--momentum $p_\mu$, expressed in the 
coordinate system $x^\mu$, lies in $V$. The definition of $f$ implies that
\begin{eqnarray}
 N_{W\times V} = \int_W \dd^3 x^i \int_V \dd_4 v_\nu \,f(x^\mu,v_\nu)\,\frac{v^0}{m}. \label{eqfdef2}
\end{eqnarray}
This expression may serve as an alternative definition of $f$.

We may reduce the number of variables in (\ref{eqEinsteinVlasov1}) by integrating $f$ partially over $m = \sqrt{-v^\alpha \,v_\alpha}$. This is possible because the Einstein--Vlasov
equations are insensitive to masses of individual particles in the following sense: two particles of mass $m$ at the same
point of spacetime and the same initial velocity follow the same world--line and generate the same gravitational field
as one particle of mass $2m$ with the same initial condition. Therefore we can replace any collection particles of different masses with a collection of particles of the same fixed mass, generating the same gravitational field
and evolving in the same manner as the original one. In the standard treatment of the subject it is often assumed from the beginning that  
the four--momenta of all particles lie on the same mass shell, i. e. 
\begin{eqnarray}
 f(x^\mu, v_\nu) = \delta(v_\mu\,v_\nu\,g^{\mu\nu} + m^2)\, \tilde f(x^\mu, v^j)
\end{eqnarray}
and the equations are written down in terms of $\tilde f$ \cite{ehlers:1973, andreasson:2011LRR....14....4A}. While this leads to reducing the number of variables by one, we have found that from the computational point of view it is slightly simpler to allow particles of different masses. We will consider therefore the distribution function depending on all four 
components of four--momentum. This is nevertheless entirely equivalent to the single mass shell formulation.

In order to ensure the convergence of (\ref{eqEinsteinVlasov1}) we need to
impose suitable fall--off conditions for $f$ at $v \to \infty$
\cite{andreasson:2011LRR....14....4A} .
It suffices to assume that in each point $x \in \mm$ $f$ vanishes for sufficiently large $p$.

\section{The geometric setting for the Einstein--Vlasov matter} \label{sec:EV-geometric}

As in \cite{kijowski-1998}, we begin by introducing the abstract material space, denoted here by $\bb$.
By assumption the particles in our case can be distinguished by labeling which involves the position of particle at a given instant of time (three values of coordinates) and the value of its four--momentum. The material space as a manifold must therefore be seven--dimensional. It should be equipped with a positive measure $\epsilon$, whose integral yields the total number of particles contained in a domain of $\bb$, and a positive, smooth function $\mu$, which gives the mass of particles contained
in an infinitesimal neighbourhood of a given point in $\bb$. Altogether, the material space is defined as the triple $(\bb, \epsilon, \mu)$. For further convenience, we introduce also a coordinate system $(\xi^A)$, $A = 1..7$, on $\bb$. 

Let $I$ denote a fixed time interval $t_0\le t \le t_1$ and let $\zz$ be the Cartesian product $\zz = I\times \bb$. 
By $\bb_t$ we will denote the constant time section of $\zz$, isomorphic to $\bb$.
The complete time evolution of matter degrees of freedom can be now encoded in a single mapping 
\begin{eqnarray}
\ff : \zz \to P^*_+.
\end{eqnarray}
satisfying a number of constraint equations.

First, for the sake of convenience we assume that the time coordinate $t$ on $\zz$ agrees with 
$x^0$ on $P^*_+$
\begin{eqnarray}
 t = x^0\circ \ff. \label{eqconstraint1}
\end{eqnarray}
This is a useful gauge choice which ensures that $\bb_t$ is mapped into $\Sigma_t$ for every $t$.
We will denote the restriction of $\ff$ to $\bb_t$ by $\ff_t$.
With the help of coordinate systems the full mapping can now be described by seven functions $x^i(t,\xi^A)$ and $v_\nu(t,\xi^A)$.

Secondly, we must require that the world--lines of particles, i.e. the images of curves $\xi^A = \const$, are four--momentum lifts. From (\ref{eq4mlift}) this amounts to
\begin{eqnarray}
  v_\nu(t,\xi^A) = \frac{\mu\,\dot x^\mu\,g_{\mu\nu}}{\sqrt{-\dot x^\rho\,\dot x^\sigma\,g_{\rho\sigma}}} \label{eqconstraint2} 
\end{eqnarray}
where $\dot x^\mu = \pard{x^\mu}{t}$.
A simple consequence of this equation is that the mass function $\mu$ coincides with the pullback of $m = \sqrt{-v^\alpha\,v_\alpha}$, so
the length of the four--momentum of each particle is indeed equal to $\mu$ and conserved during the motion.

\begin{figure}
 \includegraphics[scale=0.8]{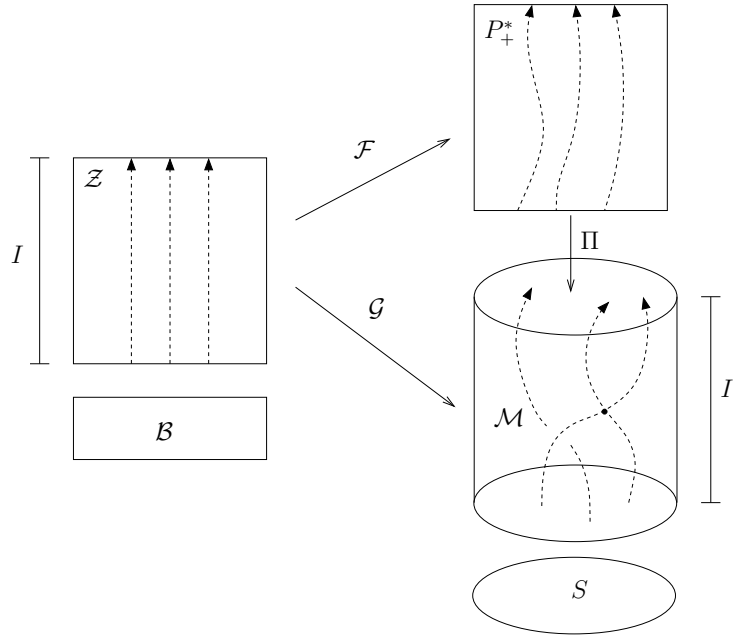}
 \caption{Geometric description of the Vlasov matter. Manifold $\zz$, Cartesian product of the material space $\bb$ and time interval $I$, is ruled by $\xi^A= \const$ curves. These curves are mapped by $\ff$ to the lifts of individual particle trajectories to $P^*_+$. These lifts do not intersect, although the corresponding world--lines on $\mm$ do.}
\label{fig-mappings}
\end{figure}

The second condition implies also that $\ff$ can be unambiguously reconstructed from its projection $\Gg$
down to $\mm$ via the bundle projection $\Pi$:
\begin{eqnarray}
 \Gg = \Pi\circ \ff
\end{eqnarray}
(see Fig. \ref{fig-mappings}).

Indeed, $\Gg$ encodes only the individual trajectories of the single particles $x^i(t,\xi^A)$, but together 
with the mass function that is enough to calculate the four--momentum at any time and calculate the four--momentum lift to $P^*_+$. A pleasant consequence of this
observation is that it is possible to eliminate $v_\nu(t,\xi^A)$ as variables and encode the momentary
state of the system in three fields $x^i(t,\xi^A)$, giving for a fixed $t$ the restriction $\ff_t$ of $\ff$.

\subsection{Connection to the standard formalism}
We will now establish the connection of the formalism described above to the standard description via the distribution function. 
We define the particle current on $\zz$ as a densitized vector field $\tilde J$, given in the introduced coordinate system $(t, \xi^A)$ as
$\tilde J^t = \phi(\xi^A)$, where $\epsilon = \phi(\xi^A)\,\dd \xi^1\wedge\dots\wedge\dd\xi^7$, and all other components vanish. $\tilde J$ yields
the number of particles intersecting any oriented hypersurface in $\zz$ when integrated over it. It is the vector density counterpart of the scalar density $\epsilon$
on $\zz$ \cite{KIJOWSKI1992207, PhysRevD.41.1875}. We assume $\epsilon$ to have compact support in order to make sure that the integrals in (17) converge. 

Since $\ff$ is a diffeomorphism, it is invertible and all geometric objects can be pushed back and forth between $\zz$ to $P^*_+$. The pushforward of the particle current to $P^*_+$ by $\ff$ will be denoted by $J$. It is tangent to the trajectories of single particles and, just like
 $\tilde J$, it encodes the flux of particles: its integral yields the number of particles crossing a particular hypersurface in $P^*_+$. The particle trajectories generated by $J$ must satisfy (\ref{eqconstraint2}), therefore the horizontal part of $J$, by definition proportional to $\dot x^\mu$, must also be proportional to $v_\nu$:
\begin{eqnarray}
 J_x^\mu = C\,v_\nu\,g^{\mu\nu}. \label{eqpppp1}
\end{eqnarray}
The proportionality constant $C$ contains the information about the particles density and can be linked to the distribution function $f$ in the following way: 
for a given hypersurface $D \in \Sigma_t$ the number of particles crossing it is
$N_D = \int_D \dd x^1\,\dd x^2\,\dd x^3\,\dd v_0\,\dots\dd v_3\,J_x^0.$
A comparison with (\ref{eqfdef2}) leads immediately to 
\begin{eqnarray}
 C = \frac{f(x^\mu, v_\mu)}{m}. \label{eqpppp2}
\end{eqnarray}

It is useful to write equations (\ref{eqpppp1}) and (\ref{eqpppp2}) in a geometric, coordinate--invariant way. 
As a vector density, the particle current can be decomposed into a product of a given density and a vector field parallel to $J$. 
As the density we take $\kappa$, the invariant volume form on $P^*_+$. This gives rise to decomposition of the form
\begin{eqnarray}
 J = \frac{f\,\kappa}{m}\,V \label{eq:JfkappamV}
\end{eqnarray}
where the horizontal part of $V$ satisfies
\begin{eqnarray}
 V_x^\mu = g^{\mu\nu}\,v_\nu.
\end{eqnarray}

Applying the equation for pushforward to $\tilde J$ we can obtain the equation for $J$, and thus also for $f$, expressed in terms of $x^i(\xi^A,t)$ and $\phi(\xi^A)$: 
\begin{eqnarray}
 f = \frac{m\,\phi\Big(\xi^A(t,x^i,v_\nu)\Big)}{v^0}\,|D\ff_t|^{-1}\,\label{eqdeff}
\end{eqnarray}
where $|D\ff_t|$ is the absolute value of the Jacobian of $\ff_t$
\begin{eqnarray}
 |D\ff_t| = \left|\det\left(\pard{(x^i,v_\nu)}{(\xi^A)}\right)\right|
\end{eqnarray}
(note that $\phi(\xi^A)$ is composed with the inverse of $\ff_t$ in the formula above).

It is easy to prove that $f$ is subject to a conservation equation of the Vlasov type. Note that $\tilde J$ is by definition divergence--free and preserves the mass function $\mu$
\begin{eqnarray*}
 \dyv \tilde J &=& 0 \\
 \tilde J (\mu) &=& 0.
\end{eqnarray*}
These properties must be shared by $J$ irrespective of the mapping $\ff$, i.e.
\begin{eqnarray}
 \dyv J &=& 0 \label{eqdivj}\\
 J(m) &=& 0. \label{eqJm}
\end{eqnarray}
Hence
\begin{eqnarray}
 0 = \dyv J = \Lie_V\left(\frac{f}{m}\,\kappa\right) = \frac{\kappa}{m}\,\Lie_V(f\,\kappa) 
\end{eqnarray}
or
\begin{eqnarray}
 \kappa\,V(f) + f\,\Lie_V \kappa = 0.
\end{eqnarray}
So far we did not assume anything about the particle world--lines themselves, but if we  assume that the particles follow geodesics, we have $V = G$ and consequently
\begin{eqnarray}
 G(f)=0. 
\end{eqnarray}

Define now the vector field $U = \frac{G}{m}$. $U$ is  the normalized
counterpart of $G$, so $U_x^\mu$ gives the 4-velocity of a particle instead of its 4-momentum.
We see that (\ref{eq:JfkappamV}) is then equivalent to 
\begin{eqnarray}
J = f\,\kappa\,U,
\end{eqnarray}
in full analogy to the standard dust or fluid \cite{KIJOWSKI1992207, PhysRevD.41.1875}.

For the purpose of this paper we introduce also the particle mass current
\begin{eqnarray}
 M = m\,J, \label{eqMdef}
\end{eqnarray}
which measures the total invariant mass carried by particles through a given hypersurface. 
It shares properties (\ref{eqdivj}--\ref{eqJm}) with $J$, and its horizontal part is given by
\begin{eqnarray}
 M_x^\mu = f\,v_\nu\,g^{\mu\nu}. \label{eqMx}
\end{eqnarray}

\section{The Lagrangian variational principle}\label{sec:lagr} 

Let $\AAA \subset \mm$  be a compact domain, topologically $C \times \left[t_0,t_1\right]$. We will assume that the basic variables, the metric and the mapping $\Gg$, are fixed on the boundary $\partial \AAA= C_{t_0} \cup C_{t_1} \cup (\partial C \times \left[t_0, t_1\right])$.

In order to describe $\Gg$ using 
scalar functions, we consider the pullbacks of spacetime coordinates to $\zz$: 
\begin{equation}
X^\mu = x^\mu \circ \Gg. 
\end{equation}
With gauge condition (\ref{eqconstraint1}) satisfied we have $X^0(t,\xi^A) = t$, so the matter degrees of freedom are completely described by three scalar functions $X^i(t,\xi^A)$.

In an Einstein--Vlasov system all particles should follow geodesic world--lines. We will therefore take 
for the matter action functional the integral of the standard single--particle action, whose variation yields geodesic equations, over all particles. There are many choices for single--particle action for geodesics, but we shall use exclusively
\begin{eqnarray}
 S_{1p} &=& \int_{s_0}^{s_1} L_{1p}\left(\frac{\dd x^\mu}{\dd s}, x^\nu\right)\,\dd s \\
 L_{1p}\left( \frac{\dd x^\mu}{\dd s}, x^\nu\right) &=& -m\,\sqrt{-g_{\mu\nu}(x^\alpha)\,\frac{\dd x^\mu}{\dd s}\,
\frac{\dd x^\nu}{\dd s}}. \label{eqHomLag}
\end{eqnarray}
where $x^\mu(s)$ is a space--like  world--line. 
This particular Lagrangian is homogeneous in the velocities, i.e.
\begin{eqnarray}
  L_{1p}\left(\lambda \, \frac{\dd x^\mu}{\dd s}, x^\nu\right) = |\lambda|\,L\left( \frac{\dd x^\mu}{\dd s}, x^\nu\right) \label{eqHomogenL}
\end{eqnarray}
This fact has an interesting consequence that the action is invariant with respect to reparametrization of the world--line: for any timelike world--line $x^\mu(s)$ and any 
strictly increasing and $C^1$ function $\tau(s)$ we have
\begin{eqnarray}
 S_{1p} = -\int_{s_0}^{s_1} m \sqrt{-g_{\mu\nu} \frac{\dd x^\mu}{\dd s} \frac{\dd x^\nu}{\dd s}} \dd s = -\int_{\tau(s_0)}^{\tau(s_1)} m\sqrt{
-g_{\mu\nu} \frac{\dd x^\mu}{\dd \tau} \frac{\dd x^\nu}{\dd \tau}} \dd\tau \label{eqHomAct}.
\end{eqnarray}
 Thus if a curve $x^\mu(s)$ minimizes the action, it does so irrespective of parametrization. 
Assuming that the variation of $\delta x^\mu$ vanish at the endpoints of integration, the variation of the action reads
\begin{eqnarray}
 \delta S_{1p} = -\int_{s_0}^{s_1} \left(\frac{\dd ^2 x^\mu}{\dd s^2} + \Gamma\UDD{\mu}{\nu\sigma}\,\frac{\dd x^\nu}{\dd s}\,\frac{\dd x^\sigma}{\dd s} - \lambda\,\frac{\dd x^\mu}{\dd s}\right)\,\delta x^\mu  \label{eqGeod}
\end{eqnarray}
where 
\begin{eqnarray}
 \lambda = \frac{1}{2}\,\frac{\dd}{\dd s} \ln \left|g_{\mu\nu}\,\frac{\dd x^\mu}{\dd s}\,\frac{\dd x^\nu}{\dd s}\right|.
\end{eqnarray}
Thus $\delta S_{1p} = 0$ implies the equation for a geodesic with non--affine, arbitrary parametrization 
\begin{eqnarray}
 \nabla_{\dot x} \dot x = \lambda\,\dot x.
\end{eqnarray}
This conclusion also holds if we fix one of the coordinates, for example by requiring that $x^0 = s$.

Let us now return to the matter action functional.
The integral over $\Gg^{-1}(\AAA)$ of the single--particle action (\ref{eqHomAct}) reads
\begin{eqnarray}
 S_m &=& \int_\bb \epsilon \int_{\tau_0}^{\tau_1} L_{1p}\big(X^\mu,\dot X^\mu,g_{\mu\nu}(X^\alpha)\big)\,\dd t  \nonumber\\
 L_{1p} &=& - 
\mu\sqrt{-g_{00}(X^\alpha) - 2g_{0i}(X^\alpha)\,\dot X^i - g_{ij}(X^\alpha)\,\dot X^i\,\dot X^j} \label{eqSm}. 
\end{eqnarray}
Dot denotes here partial differentiation with respect to $X^0$ ($\frac{\partial}{\partial t}$) keeping the $\xi^A$'s constant, thus by definition $\dot X^0 = 1$. 
$\tau_0 = \tau_0(\xi^A)$ and $\tau_1 = \tau_1(\xi^A)$ denote the endpoints of intersections of $\xi^A = \const$ curves and $\Gg^{-1}(\AAA)$.
%For computational convenience we have fixed the curve parametrization to be given by $x^0$. This does not of course affect the equations of motion.

We will now demonstrate that  vanishing of variations with respect to $X^i$ leads to the geodesic equations for
individual particles, while the variation with respect to $g_{\mu\nu}$ gives exactly expression (\ref{eqEinsteinVlasov1}) for the stress energy tensor.

\subsection{Variations with respect to $X^i$}

Consider variations of the type $\delta g_{\mu\nu} = 0$, $\delta X^i \neq 0$. 
The variation of $S_m$ reads 
\begin{eqnarray}
 \delta S_m = -\int_\bb \epsilon\, \delta\left( \int_{\tau_0}^{\tau_1} \mu(\xi^A) \sqrt{-\dot X^\mu\,\dot X^\nu\,g_{\mu\nu}}\,\dd t\right). 
\end{eqnarray}
The variation is the integral of independent variations of single--particle lagrangians and each of them
must vanish if $\delta S_m$ vanishes. By direct calculations one may check that it leads to geodesic equations of the form of
(\ref{eqGeod}), i.e.
\begin{eqnarray}
 && \ddot X^\mu + \Gamma^\mu_{\phantom{\mu}\nu\sigma}\,\dot X^\nu\,\dot X^\sigma = \lambda\,\dot X^\mu \nonumber \\
 && \lambda = \frac{1}{2}\,\frac{\dd}{\dd t}\,\ln\left| \dot X^\alpha\,\dot X^\beta\,g_{\alpha\beta}\right| \label{eqGeod2}
\end{eqnarray}
with the gauge condition $X^0 = t$ fixing the parametrization. These equations determine the mapping $\Gg$ uniquely and thus, 
as we have shown at the end of section \ref{sec:EV-geometric}, they fix $\ff$ and imply that the distribution function on $P^*_+$ satisfies the Vlasov equation (\ref{eqGf}).

\subsection{Variation with respect to the metric}

The metric tensor depends explicitly on the spatial variables $g_{\mu\nu} = g_{\mu\nu}(x^\alpha)$. Varying $S_m$ with respect to the metric components ($\delta g_{\mu\nu} \neq 0$) while keeping all values of $X^i(t,\xi^A)$ fixed yields 
\begin{eqnarray}
 \delta S_m = \frac{1}{2} \int_\bb \epsilon\left(\int_{\tau_0}^{\tau_1}\frac{m\,\dot X^\mu \,\dot X^\nu \,\delta g_{\mu\nu}}{\sqrt{-\dot 
X^\alpha\,\dot X_\alpha}}\,\dd t\right) \label{eqvargSm1}
\end{eqnarray}
We can use $\ff$ to perform the integration over $\Pi^{-1}(\AAA)$ instead of $\Gg^{-1}(\AAA)$:
\begin{eqnarray}
 \delta S_m &=& \frac{1}{2}\int_{\Pi^{-1}(\AAA)} \frac{m\,\dot X^\mu \,\dot X^\nu \,\delta g_{\mu\nu}}{\sqrt{-\dot 
X^\alpha\,\dot X_\alpha}} \left|D\ff_t\right|^{-1}\,\phi\big(\xi^A(x^\alpha,v_\beta)\big) \,\dd^4 x^\mu\,\dd^4 v_\nu \\
 &=& \frac{1}{2}\int_{\Pi^{-1}(\AAA)} \frac{m\,f\,\dot X^\mu\,\dot X^\nu\,\delta g_{\mu\nu}}{\left(-\dot X^\alpha\,\dot X_\alpha\right)} \,\dd^4 x^\mu\,\dd^4 v_\nu
\end{eqnarray}
(we have used equation (\ref{eqdeff}) to get rid of the inverse Jacobian). Now using (\ref{eqconstraint2}) we finally get
\begin{eqnarray}
 \delta S_m &=& \frac{1}{2}\int_{\Pi^{-1}(\AAA)} \frac{f\,v^\mu\,v^\nu}{m}\,\delta g_{\mu\nu}\,\dd^4 x^\mu\,\dd^4 v_\nu.
\end{eqnarray}
From (\ref{eqQQQ1}) we conclude that the stress--energy tensor at a given point $p$ is equal to
\begin{eqnarray}
 T^{\mu\nu} = \int_{\Pi^{-1}(p)} \frac{f\,v^\mu\,v^\nu}{m\,\sqrt{-g}}\,\dd^4 v_\alpha,
\end{eqnarray}
in full agreement with (\ref{eqEinsteinVlasov1}).

\subsection{Hamiltonian}

It is possible to derive the Hamiltonian of the Einstein--Vlasov system directly from (\ref{eqSm}).
The momentum fields $p_i(t, \xi^A)$ conjugate to the fields $X^i(t, \xi^A)$ are given by
\begin{eqnarray}
 p_i = \frac{\partial L}{\partial \dot X^i} = \frac{\mu\, g_{\mu i}\,\dot X^\mu}{\sqrt{-\dot X^\mu\,\dot X^\nu\,g_{\mu\nu}}},
\end{eqnarray}
i.e. they are the lower--index space components of the standard single particle four--momentum.

By performing the Legendre transform between
$p_i$ and $\dot X^i$ we obtain the Hamiltonian ${\cal H}_m$, which turns out to be
simply the integral of the time component of the particle four--momentum:
\begin{eqnarray}
 {\cal H}_m = \int_\bb \left(p_i \dot X^i - L\right)\epsilon = 
-\int_\bb p_0 \,\epsilon, \label{eqHm}
\end{eqnarray}
where $p_0$ is expressed in terms of $p_i$ and $X^\mu$.
\begin{eqnarray}
p_0 &\equiv& p_0(p_i,X^\mu) = \frac{p^0}{g^{00}} - \frac{p_i\, g^{0i}}{g^{00}} \label{eqp0}\\
p^0 &\equiv& p^0(p_i,X^\mu) = \sqrt{\left(p_i\,g^{0i}\right)^2 - \left(\mu^2 + p_k\, p_l \,g^{kl}\right) g^{00}} \\
g^{\mu\nu} &\equiv& g^{\mu\nu}(X^\alpha).
\end{eqnarray}
If we use the ADM variables
$h_{ij},N^i,N$ instead of the metric components $g_{\mu\nu}$
\begin{eqnarray}
 g_{\mu\nu} = \left(
\begin{array}{rr}
-N^2 + N^k\,N^l\,h_{kl} & N^k\,h_{kn} \\
N^k\,h_{km} & h_{mn}  
\end{array} \right),
\end{eqnarray}
 the Hamiltonian turns out to consist of familiar scalar and vector terms
\begin{eqnarray}
 {\cal H}_m = \int_\bb \left(-p_i\,N^i + N\,\sqrt{\mu^2 + p_k\,p_l\,h^{kl}}\right)\,\epsilon.
\end{eqnarray}
These terms enter the vector and scalar constraints in the ADM formulation and the total Hamiltionian of the Vlasov matter coupled with gravitational field reads 
\begin{eqnarray}
 {\cal H}_\textrm{tot} = \frac{1}{16\pi G}\,\int \dd^3 x^i\,\sqrt{h} \,\left( N\,\left({\cal H} + 16\pi G\,\sqrt{h}\,T^{00}\right) + N_i\,\left({\cal H}^i + 16\pi G\,\sqrt{h}\,T^{0i}\right)\right)
\end{eqnarray}
where ${\cal H}$ and ${\cal H}^i$ are the scalar and vector constraints. 

Finally note that by the virtue of (\ref{eqHm}), (\ref{eqdeff}) and (\ref{eqEinsteinVlasov1}) the value Hamiltonian is the integral of the null--null component of the stress--energy tensor over the constant coordinate time hypersurface
\begin{eqnarray}
 {\cal H}_m &=& -\int  T\UD{0}{0}\,\sqrt{-g}\,\dd^3 x^i .
\end{eqnarray}
In this form it was used to investigate the stability of stationary Einstein--Vlasov equation solutions [Kandrup].

\section{Eulerian variational principle} \label{sec:eul} 

We will now present the same variational principle from a different, Eulerian point of view.
The evolution of the Vlasov matter will be described in terms of flux of particles in covector future light cones bundle $P_+^*$. 
 In other words, we can understand the Vlasov matter
as dust moving in the tangent bundle rather than in the spacetime. The matter main variable will be the particle mass current vector density $M$ defined in
(\ref{eqMdef}).

Dust is most commonly described by the its velocity field and density, but we argue here that the most natural geometric objects to measure the current of particles are vector densities. This is because they simultaneously carry information about
the direction of movement of a single particle and the particle density at a given point. In other words, they combine
the density and the velocity into a single object. Indeed, a vector density can be
understood geometrically as a product of a density and an ordinary vector field. In appendix $\ref{app}$ we briefly remind some of its main features.

\subsection{The Eulerian variational principle}

We now spell out the variational principle in the language of mass current. The state of matter is described by a vector density $M$ on $\Pi^{-1}(\AAA) \subset P_+^*$,  while the gravitational degrees of freedom are described by the metric tensor on $\AAA$ (and connection if we consider Palatini--type action). The matter action,
considered now as a functional of the metric $g_{\mu\nu}$ and $M$, is given by 
\begin{eqnarray}
 S_m = -\int_{\Pi^{-1}(\AAA)} \sqrt{-g_{\mu\nu}\,M_x^\mu\,M_x^\nu}\,\dd^4 x^\alpha\,\dd^4 v_\beta \label{eqactionM}.
\end{eqnarray}
This action is merely (\ref{eqSm}) rewritten in terms of $M$ and integrated over $\Pi^{-1}(\AAA)$ rather than $\zz$.
It is possible to express it in a coordinate--independent way
\begin{eqnarray}
 S_m = - \int_{\Pi^{-1}(\AAA)} \sqrt{-\Pi^*g\left(M,M\right)}.
\end{eqnarray}
Note that the integrand is a scalar density, so the integral does not require a measure to be specified.

Mass current $M$ is not entirely arbitrary, but subject to constraints. We will now spell them out one by one.
First, we assume that $M$ is divergence--free
\begin{eqnarray}
 \dyv M &=& 0. \label{eqcnsdivM}
\end{eqnarray}
It should also be tangent to the mass shells so that every integral curve remained at its mass shell
\begin{eqnarray}
\left(M_x^\mu\,\frac{\partial }{\partial x^\mu} + M_{v\,\nu}\,\frac{\partial }{\partial v_\nu}\right)\,\left(v_\alpha\,v_\beta\,g^{\alpha\beta}\right) = 0.
\label{eqcnsmass}
\end{eqnarray}
These conditions together imply that the number of particles as well as the mass of each individual particle is conserved during the motion.
Finaly we assume that the integral curves of $M$ are lifts of  particle world--lines, which means that the horizontal components of $M$ are proportional to $v^\mu$, i.e. satisfying (\ref{eqMx}).

The constraints above are all kinematic constraints -- we only take into account vector densities which satisfy them from the beginning.

\subsection{Variation with respect to the metric}

Now we move on to calculate the variation of $S_m$ and derive the equations of motion.
First we consider variations of the type $\delta g_{\mu\nu} \neq 0$, $\delta M = 0$, with $\delta g_{\mu\nu}$ vanishing at $\partial \AAA$.
The calculation is straightforward; by the virtue of (\ref{eqQQQ1}) the stress--energy tensor takes of the form
\begin{eqnarray}
 T^{\mu\nu} = (-g)^{-1/2} \int \frac{M_x^\mu\,M_x^\nu}{\sqrt{-g_{\alpha\beta} \,M_x^\alpha\,M_x^\beta }}\,\dd^4 v_\nu.
\end{eqnarray}
After substituting (\ref{eqMx}) we obtain the correct expression (\ref{eqEinsteinVlasov1}) for the stress--energy tensor.

\subsection{Variation with respect to $M$}

In contrast to unconstrained variations of $g_{\mu\nu}$, variations of $M$ are subject to a number of constraints.
Firstly, 
in the spirit of
\cite{taub-1954} we only consider
variations which arise from displacing the individual particles' world--lines. The displacement is described by a vector field $X$ on $P_+^*$, vanishing of the boundary of the integration domain $\Pi^{-1}(\AAA)$. The variations take the form of
\begin{eqnarray} 
 \delta g &=& 0 \\ 
 \delta M &=& \Lie_X M \label{eqdeltaM}.
\end{eqnarray}
In principle $X$ may be arbitrary, but the  consistency of (\ref{eqdeltaM}) with (\ref{eqMx}) and (\ref{eqcnsdivM}--\ref{eqcnsmass}) imposes restrictions on admissible vector fields $X$.
By simple computation we can check that (\ref{eqcnsdivM}) is preserved by all variations
of type (\ref{eqdeltaM}), but  conditions (\ref{eqcnsmass}) and (\ref{eqMx}) do impose restrictions on $X$: (\ref{eqMx}) is preserved by variations if there exists a function $c$ such that
\begin{eqnarray}
 (\delta_X M)_x^\mu = c\,v_\nu\,g^{\mu\nu} \label{eqcnsliftinf}.
\end{eqnarray}
With the help of (\ref{eqLieK}) and (\ref{eqMx}) we can put (\ref{eqcnsliftinf}) in the following form
\begin{eqnarray}
X_v^\mu = \tilde c\,v^\mu - X_x^\alpha\,\frac{\partial g^{\mu\nu}}{\partial x^\alpha}\,v_\nu + v^\alpha\,\frac{\partial X_x^\mu}{\partial x^\alpha} + 
f^{-1}\,M_{v\,\alpha}\,\frac{\partial X_x^\mu}{\partial v_\alpha} \label{eqXv}.
\end{eqnarray}
where $\tilde c$ is again an arbitrary function.
Thus the vertical part of $X$ is determined up to a multiple of $v^\mu$  by the horizontal part $X_x^\mu$, which in turn is completely free.  
The value of $\tilde c$ is fixed by condition (\ref{eqcnsmass}). Recall that it simply asserts that the integral curves
of $M$ lie on the mass shells. Since we need the curves to remain at the mass shell after the variation, we
must require that $X$ is tangent them as well
\begin{eqnarray}
 X(-g^{\mu\nu}\,v_\mu\,v_\nu) = 0
\end{eqnarray}
or
\begin{eqnarray}
  X_{v\,\mu} g^{\mu\nu}\,v_\nu =-\frac{1}{2}\, X_x^\alpha \,\frac{\partial g^{\mu\nu}}{\partial x^\alpha}\,v_\mu\,v_\nu. \label{eqXvv}
\end{eqnarray}
Clearly the second formula, multiplied by $v_\mu$, fixes the value of $\tilde c$ everywhere.
Summarizing, admissible variations come from vector fields $X$ in which the vertical part is determined by the horizontal part via (\ref{eqXv}) and (\ref{eqXvv}).  

We are now ready to evaluate $\delta S_m$. In order to facilitate the calculation, we prove a more general statement about homogeneous Lagrangians and actions
of type (\ref{eqactionM}):
\begin{theorem}
 \label{thm1}
Given a lagrangian $L(\dot x^i, x^j)$ ($i,j = 1\dots n$) homogeneous in the dotted variables. Let $K$ be a vector density field on a
domain $E \subset {\bf{R}}^n$ of vanishing divergence and
let 
\begin{eqnarray}
S = \int_E L(K^i(x^j), x^j) \label{eqSSS}
\end{eqnarray}
be the action. Consider variations of $K$ arising from integral curves displacements
\begin{eqnarray}
 \delta_X K = \Lie_X K, \label{eqVarj}
\end{eqnarray}
$X$ being a vector field in $D$. Assume also that at every $p \in \partial D$ either $X^i = 0$ or $K_j = 0$.
Then the variation of (\ref{eqSSS}) reads
\begin{eqnarray}
 \delta_X S = \int_E  \left(K^i\,\partial_i K^k \,\frac{\partial^2 L}{\partial \dot x^l \partial \dot x^k} + K^i\,\frac{\partial^2 L}{\partial \dot x^l \partial x^i} - \frac{\partial L}{\partial x^l}\right) X^l
\dd^n x \label{eqVarS}
\end{eqnarray}
where in all derivatives of $L$ we substitute $K^l(x^k)$ for $\dot x^l$.
\end{theorem}
For the proof see the Appendix.

Recall that the variation of the single particle action
\begin{eqnarray}
 S_{1p} = \int_{t_0}^{t_1} 
L(\dot x^i(t), x^j(t)) \,\dd t \label{eqSsingle}
\end{eqnarray}
has the form of
\begin{eqnarray}
 \delta S_{1p} = -\int_{t_0}^{t_1} \left(\ddot x^k(t) \,\frac{\partial^2 L}{\partial \dot x^l \partial \dot x^k} + \dot x^i(t)\,\frac{\partial^2 L}{\partial \dot x^l \partial x^i} - \frac{\partial L}{\partial x^l}\right) \,\delta x^l(t) \,\dd t \label{eqVarSsingle}
\end{eqnarray}
provided that $\delta x^l(t)$ vanishes at $t = t_0$ and $t_1$. By comparing (\ref{eqVarS}) and (\ref{eqVarSsingle})
we clearly see that the world--line displacement variational principle simply states that all curves of the congruence
generated by $K^l$ are stationary points of the single particle action. Being a vector \emph{density}, $K^l$ does not specify the parametrization of these curves, but (\ref{eqSsingle}) itself is insensitive to parametrization too since $L$ is homogeneous.

We now apply the theorem to (\ref{eqactionM}). The single--particle lagrangian reads
\begin{eqnarray}
L_{1p}(\dot x^\mu,\dot v_\nu,x^\mu,v_\nu) = -m\sqrt{-\dot x^\mu\,
\dot x^\nu\,g_{\mu\nu}(x^\alpha)}.
\end{eqnarray}
 When viewed as a function on $P_+^*$, it does not involve $v_\nu$ or $\dot v_\nu$ and therefore the action is insensitive to the vertical coordinates $v_\mu$. This means that only the horizontal coordinates $X^\mu_x$  appear in (\ref{eqVarS}) if we apply Theorem \ref{thm1}. This is quite fortunate, 
since these are exactly the coordinates of $X$ which are free, in contrast to $X_{v\,\nu}$ which, due to the constraints, are not. Since all appearing coordinates of $X$ are free, we immediately conclude that
the vanishing of $\delta S$ under the admissible variations implies the vanishing of the four remaining terms in brackets in (\ref{eqVarS}). 
 
From Theorem \ref{thm1} we know that it suffices to substitute the ordinary derivatives in the single particle Euler--Lagrange equations (\ref{eqGeod}) 
by appropriate coordinates and derivatives of $M$ to get the four equations of motion for the action (\ref{eqactionM}):
\begin{eqnarray}
 && M_x^\nu \, \pard{M_x^\mu}{x^\nu} + M_{v\,\nu}\,\pard{M_x^\mu}{v_{\nu}}+
 \Gamma\UDD{\mu}{\nu}{\sigma}\,M_x^\nu\,M_x^\sigma = \nonumber\\
&=& \frac{1}{2}\,\left(M_{x}^\nu\,\pard{\ln \left|g_{\gamma\delta} M_x^\gamma\,M_x^\delta\right|}{x^\nu} + M_{v\,\nu}\,\pard{\ln 
\left|g_{\gamma\delta} M_x^\gamma\,M_x^\delta\right|}{v_\nu}\right) M_x^\mu. \label{eqMMM} 
\end{eqnarray}
Note that these four equations simply imply that the curves generated by $M$ are geodesic if projected down to $\mm$.  
Together with the constraints (\ref{eqcnsdivM})--(\ref{eqcnsmass}) and (\ref{eqMx}) they fix $M$ completely by stating that the integral curves are lifts of
geodesics to the tangent bundle. We will show this now in details.

We now substitute (\ref{eqMx}) to (\ref{eqMMM}) and  get
\begin{eqnarray}
&& f\,v^\nu\, \pard{}{x^\nu}(f\,v_\sigma\,g^{\sigma\mu}) + M_{v\,\nu}\,\pard{}{v_\nu}(f\,v_\sigma\,g^{\sigma\mu}) + 
f^2\,\Gamma\UDD{\mu}{\rho}{\sigma}\,v^\rho\,v^\sigma = \nonumber\\
&=& \frac{1}{2}\left(f\,v^\nu\pard{}{x^\nu}\,\ln \left|f^2\,g^{\rho\sigma}\,v_\rho\,v_\sigma\right| + 
M_{v\,\nu}\,\pard{}{v_\nu}\,\ln |f^2\,g^{\rho\sigma}\,v_\rho\,v_\sigma|\right)\,f\,v^\mu.
\end{eqnarray}
We simplify this equation using the identity
\begin{eqnarray}
 \pard{}{x^\nu} g^{\alpha\beta} + \Gamma\UD{\alpha}{\kappa\nu}\,g^{\kappa\beta} + \Gamma\UD{\beta}{\kappa\nu}\,
g^{\kappa\alpha} = 0 \label{eqinvg}
\end{eqnarray}
and obtain
\begin{eqnarray}
 M_{v\,\mu} - f\,\Gamma\UD{\sigma}{\mu\rho}\,v_\sigma\,v^\nu &=&
-f\,v_\mu\,\Gamma\UD{\rho}{\sigma\nu}\,\frac{v_\rho\,v^\sigma\,v^\nu}{v^\alpha\,v_\alpha} +
f\,v_\mu\,\frac{v^\nu\,M_{v\,\nu}}{v^\alpha\,v_\alpha}
\end{eqnarray}
or
\begin{eqnarray}
 M_{v\,\nu}\,\left(\delta\UD{\nu}{\mu} - \frac{v^\nu\,v_\mu}{v^\alpha\,v_\alpha}\right) = 
f\,\Gamma\UD{\sigma}{\nu\rho}\,v_\sigma\,v^\rho\,\left(\delta\UD{\nu}{\mu} - \frac{v^\nu\,v_\mu}{v^\alpha\,v_\alpha}\right). \label{eqMwithproj}
\end{eqnarray}
The expression in brackets on both sides of the equation is the orthogonal projection operator to the subspace
perpendicular to $v_\mu$. Therefore equation (\ref{eqMwithproj}) specifies $M_v$ only up to a term proportional to
$v_\mu$
\begin{eqnarray}
 M_v^\mu = f\,\Gamma\UD{\sigma}{\mu\rho}\,v_\sigma\,v^\rho + \lambda\,v_\mu. \label{eqMvperp}
\end{eqnarray}
It can be proven that  $\lambda$  vanishes due to (\ref{eqcnsmass}). Namely, 
using (\ref{eqinvg}) and (\ref{eqMx}) we can transform (\ref{eqcnsmass}) into
\begin{eqnarray}
 M_{v\,\mu}\,v^\mu = f\,\Gamma\UD{\sigma}{\mu\rho}\,v_\sigma\,v^\mu\nu^\rho
\end{eqnarray}
which, together with (\ref{eqMvperp}) yields
\begin{eqnarray}
 M_{v\,\mu} = f\,\Gamma\UD{\sigma}{\mu\rho}\,v_\sigma\,v^\rho.
\end{eqnarray}
By comparing the equation above and (\ref{eqMx}) with (\ref{eqGfield}) it is straightforward to see that the equations of motion 
 simply state that
\begin{eqnarray}
 M = f\,G\,\kappa.\label{eqMfG}
\end{eqnarray}
The Vlasov equation (\ref{eqGf}) follows then directly from the conservation equation (\ref{eqcnsdivM}).

\subsection*{Acknowledgements}
This work was completed while L.A. was in residence at Institut
Mittag-Leffler in Djursholm, Sweden, during the fall of 2019, supported by the Swedish Research Council under grant no. 2016-06596.
The authors are grateful to Hakan Andr\'easson, Marcus Ansorg, 
Gerhard Rein, and Alan Rendall for helpful discussions. 
M.K. was supported in part 
by the Foundation for Polish Science through the ''Master'' programme
and later by the by the National Science Centre, Poland (NCN) via the SONATA BIS programme, Grant No.~2016/22/E/ST9/00578 for the project
\emph{``Local relativistic perturbative framework in hydrodynamics and general relativity and its application to cosmology''}. 

\appendix 

\section{Vector densities}
\label{app}

Let $P$ be a manifold of dimension $k$ and let $K$ be a vector density on it. Let $R \subset P$ be a hypersurface in $P$ and let $TP\big|_R$ be the restriction of the tangent bundle $TP$ to $R$, i.e. $TP\big|_R = \left\{ x \in TP \mid \pi(x) \in R\right\}$, $\pi$ denotes here the projection of the tangent bundle. If the set difference 
$TP\big|_R \,\setminus\, TR$ consists of two disjoint subsets, we call the surface \emph{externally orientable}. The \emph{external orientation} is the choice of one of these subsets. A simple geometric way to single it out is to present a function $\zeta$ on a neighbourhood $R$ which vanishes on $R$ itself and whose derivative is positive on vectors from the chosen subset of $TP\big|_R$. In a less precise manner we say that the external orientation of $R$ is the choice of one side of it and function $\zeta$ is supposed to be positive on that side. 
 
Vector density $K$ can be integrated over the oriented $R$ it in the following way: take an adapted coordinate system $(\zeta, \dots, x^k)$ in which $R$ is given by the condition $\zeta = 0$ (plus perhaps inequalities imposed on other variables) and $\dd\zeta$ is consistent with the external orientation of $R$ as described in previous paragraph. The integral is by definition
\begin{eqnarray}
 \int_R  K = \int_R K^1(0,x^2,\dots, x^k)\, \dd x^2\,\dots \dd x^k. \label{eqintK}
\end{eqnarray}
 The integral is invariant in the sense that it does not depend on the coordinate system chosen, as long as $\zeta$ it satisfies the two requirements above. 

The divergence of a vector density is a scalar density given in any coordinate system by
\begin{eqnarray}
 \dyv K = \frac{\partial K^i}{\partial x^i}.
\end{eqnarray}
Alternatively the definition of divergence may take the decomposition of $K$ into the product of a scalar density and vector field as its starting point:
if $\rho$ is a scalar density and $X$ the vector field for which
\begin{eqnarray}
 K = \rho\,X \label{eqbreakup}
\end{eqnarray}
we have
\begin{eqnarray}
 \dyv K = \Lie_X\,\rho.
\end{eqnarray}
The formula above holds for any such decomposition.

The Stokes' theorem for relates the integral of divergence of $K$ over a bounded domain $Q$ in $P$ with the surface integral of $K$ over its boundary
\begin{eqnarray}
 \int_Q \dyv K = \int_{\partial Q} K
\end{eqnarray}
provided that the orientation of $\partial Q$ is given by the outward--pointing vectors.
Finally we remind here for future use the formula for the Lie derivative of $K$ with respect to a vector field $Z$
\begin{eqnarray}
 \left(\Lie_Z K\right)^i = Z^j\,\frac{\partial K^i}{\partial x^j} - K^j\,\frac{\partial Z^i}{\partial x^j} + \left(\frac{\partial Z^j}{\partial x^j}\right)\,K^i 
\label{eqLieK}
\end{eqnarray}
and the transformation law for vector densities under the change of coordinates from $(x^i)$ to $(y^j)$:
\begin{eqnarray}
 K_y^{j} = K_x^i\,\frac{\partial y^j}{\partial x^i}\,\left|\det \left(\frac{\partial x^k}{\partial y^l}\right)\right|. \label{eqtranspush} 
\end{eqnarray}
Note that the equation above may also be thought of as the equation for the push--forward of $K$ by a diffeomorphism. 

Vector densities are in many ways similar to perhaps more familiar $(k-1)$--forms. Indeed,  
if $P$ is oriented, vector densities are in one--to--one correspondence with differential forms of co-dimension $1$: namely, in any properly oriented coordinate system we can define
\begin{eqnarray}
	\alpha_{i_1\dots i_{k-1}} = \epsilon_{i_1\dots i_{k-1}i_k}\, K^{i_k}.
\end{eqnarray}
where $\epsilon$ is the Levi--Civitta symbol. The resulting differential form $\alpha$ may be integrated in a similar fashion over (internally) oriented surfaces.
Despite this equivalence, there are two reasons to prefer vector densities over differential forms in context of fluid dynamics. 

First, vector densities have a well defined direction in the following sense: we can say that $K$ is parallel to vector field $X$ if there exists a strictly positive density $\rho$ such that (\ref{eqbreakup}) holds. Thus a vector density singles out a family of vector fields differing by multiplication by a positive function. Note that these vector fields have integral curves of the same path but different parametrization. Consequently we may speak of integration curves of vector density, though, in contrast to integral curves of vector fields, they come without a preferred parametrization.  This is less obvious if we use differential forms.

The second reason is that the sign of the integrand in (\ref{eqintK}) depends only on whether the direction of $K$ is consistent with the exterior orientation of the hypersurface. This is in perfect agreement
with the definition of flux over a surface in which one should count a particle with a positive sign if it pinches through the surface in the right direction and negative otherwise. Differential forms on the other hand are only sensitive to the \emph{internal} orientation of the surface.

\section{Proof of Theorem \ref{thm1}}

We calculate the variation of (\ref{eqSSS}) under (\ref{eqVarj}). By the virtue of (\ref{eqLieK}) we have
\begin{eqnarray}
  \delta_X S &=& \int_E \dd^n x \,\frac{\partial L}{\partial \dot x^i} \,\delta_X K^i = 
\int_E \dd^n x \,\frac{\partial L}{\partial \dot x^i}\, \left(X^j\,\partial_j K^i - K^j\,\partial_j X^i + \right. \nonumber\\
&+&\left. (\partial_k X^k) K^i\right). \label{eqDeltaXI10}
\end{eqnarray}
Since the action is homogeneous in the dotted variables, it satisfies the Euler identity
\begin{eqnarray}
\frac{\partial L(\dot x^i, x^j)}{\partial \dot x^k}\,\dot x^k = L(\dot x^i, x^j),  \label{eqEulerId}
\end{eqnarray}
(it follows from (\ref{eqHomogenL}) if we differentiate it by $\lambda$), which after substituting $K^i$ for $\dot x^i$ takes the form of 
\begin{eqnarray}
 K^i\,\frac{\partial L}{\partial\dot x^i}\bigg|_{\dot x^i = K^i} = L(K^i,x^j), \label{eqEulerId2}.
\end{eqnarray}
We can plug into (\ref{eqDeltaXI10}) in the last term 
  and then perform integration by parts in the last two terms. We get
\begin{eqnarray}
 \delta_X S &=& \int_E \dd^n x \,\left(\frac{\partial L}{\partial \dot x^i}\,\partial_k K^i + (\partial_l K^l)\,
\frac{\partial L}{\partial \dot x^k} +  K^l\,\frac{\partial}{\partial x^l}\,\frac{\partial L}{\partial \dot x^k}- \frac{\partial}{\partial x^k} L\right)\,X^k + \nonumber\\
&+& \int_{\partial E} \left(L\,X^i - \frac{\partial L}{\partial \dot x^j} \,x^j\,K^i\right)\,n_i\,d^{n-1}x.
\end{eqnarray}
The surface terms vanish due to boundary conditions on $X^i$ and $K^j$. 
We finally make use of the following relations: the vanishing of divergence of the vector density $\partial_i K^i = 0$ and
\begin{eqnarray}
\frac{\partial}{\partial x^k} L\big(K^i(x^k),x^b\big) &=& \frac{\partial L}{\partial \dot x^i}\,
\partial_k K^i + \frac{\partial L}{\partial x^k} \\
\frac{\partial}{\partial x^k} \left(\frac{\partial L}{\partial\dot x^j}\bigg|_{\dot x^i = K^i}\right) &=& \frac{\partial^2 L}{\partial \dot x^j \partial \dot x^l}\,\partial_k K^l + \frac{\partial^2 L}{\partial \dot x^j \partial x^k}, 
\end{eqnarray}
and hence obtain after simplification
\begin{eqnarray}
\delta_X S &=& \int_E \dd^n x \, \left(K^k\,\partial_k K^l \,\frac{\partial^2 L}{\partial \dot x^j \partial \dot x^l} + K^k\,\frac{\partial^2 L}{\partial \dot x^j \partial x^k} - \frac{\partial L}{\partial x^j}\right) X^j.
\end{eqnarray}

\newcommand{\physrep}{Physics Reports} 

\bibliographystyle{abbrv}
\bibliography{ev-paper}

\end{document}